\documentclass[prl,amsmath,amssymb,showpacs,superscriptaddress,floatfix]{revtex4}
\usepackage{graphicx}
\usepackage{bm}
\usepackage{lineno,hyperref}
\usepackage{epstopdf}
\usepackage{epsf}
\begin{document}
\title{Dispersion of acoustic excitations in tetrahedral liquids}

\author{Yu. D. Fomin}
\affiliation{ Institute for High Pressure Physics RAS, 108840
Kaluzhskoe shosse, 14, Troitsk, Moscow, Russia}
\affiliation{Moscow Institute of Physics and Technology, 9 Institutskiy Lane, Dolgoprudny City, Moscow Region, Russia }

\date{\today}

\begin{abstract}
Investigation of the longitudinal and transverse excitations in liquids is of great importance for
understanding the fundamentals of the liquid state of matter. One of the important questions
is the temperature and density dependence of the frequency of the excitations. In our recent
works it was shown that while in simple liquids the frequency of longitudinal  excitations
increases when the temperature is increased isochorically, in water the frequency can
anomalously decrease with the temperature increase. In the present manuscript we study
the dispersion curves of longitudinal and transverse excitations of water and liquid silicon
modelled by Stillinger-Weber potential. We show that both substances demonstrate the anomaly
of the dispersion curves, but it the case of water it is more pronounced.

\end{abstract}

\pacs{61.20.Gy, 61.20.Ne, 64.60.Kw}

\maketitle

\section{Introduction}



It is well known that many properties of crystals can be
considered from the point of view of their collective excitations - phonons
\cite{solidstate,kittel}. Many properties of crystals, for instance, the heat capacity
can be efficiently described within the framework of phonon-based models. Because of this some
theories of crystalline state are indeed the theories of phonons in crystals (for instance, Debye theory).

The situation is more complex in the case of liquids.
Although the problem is in focus of many researchers now, the data on the
collective excitations in liquids are still scarce. From the point of view of
theory, there is no any simple model of liquid state (like a model of
harmonic crystal for solids) which allows to construct a zero-approach to the problem,
which can be used for further generalization. At the same time collective excitations
are of the same level of importance for the description of thermodynamic
properties of liquids like in the case of solids \cite{ropp}.

Among the most widely discusses topics in the collective excitations in liquids are the presence of
positive sound dispersion \cite{hos-fe-cu-zn,psd-ga,psd-fe,psd-sn,psd-fe-1,psd-sn-1} and the existence of transverse excitations
\cite{hos-fe-cu-zn,psd-ga,psd-fe,psd-sn,psd-fe-1,psd-sn-1}. However, until recently most of experimental works
presented measurements of the dispersion curves at a single temperature-pressure point only (usually in the vicinity of the melting
line). Because of this one could not monitor the dependence of the dispersion curves on temperature or pressure. Taking
into account that collective excitations can be used for calculation of thermodynamic properties of liquids \cite{ropp} it becomes
important to study the dependence of the excitation frequencies on temperature and pressure in order to better understand
the thermodynamics of liquids.

As shown in the previous study the frequency of longitudinal excitations in simple liquids is increased when the temperature is increased
isochorically \cite{disp-anom-135}. However, some liquids can demonstrate anomalous behavior. The most well known anomalous liquid
is water, which demonstrates dozens of anomalous properties \cite{wateranom}. We studied the spectra of collective excitations
of water (SPC/E model) and a model core-softened system (Repulsive Shoulder System (RSS)) and showed that these systems
can demonstrate anomalous behavior of the frequency of longitudinal excitations: the frequency can decrease with a temperature increase along
isochors \cite{disp-anom-wat,disp-anom-135}. At the same time the behavior of the dispersion curves of transverse excitations
is qualitatively equivalent in water and simple liquids.

From these publications we see that the liquids with anomalous behavior can demonstrate one more anomaly - the anomalous behavior of the
dispersion curves. Because of this it is very important to investigate the relation between this novel anomaly and other
anomalies in liquids. For doing this it is necessary to study the anomaly of the dispersion curves in different liquids. It is of
particular importance to find out the main trends in the behavior of the anomaly of the dispersion curves when the interaction potential
is changed.

For doing this we study the dispersion curves of liquid silicon and water along isochores which demonstrate density anomaly.
Both water and liquid silicon can be described by the same model of interaction -  Stillinger-Weber (SW) potential \cite{sw} - with different parameters.
The SW potential was parameterised for water in Ref. \cite{sw-water}. The same model was also fitted to reproduce other tetrahedral liquids,
such as carbon \cite{sw-c} and germanium \cite{sw-ge}. Importantly, these liquids are characterized by different degree of tetrahedrality, controlled by
parameter $\lambda$ (see Eq. (3) below). The tetrahedral ordering in water is stronger then in liquid silicon ($\lambda =23.15$ for water and
$21.0$ for silicon). At $\lambda$ between 18.2 and 18.6 the system crystallized into
$beta-$tin structure which allows to use it for qualitative description of tin \cite{tetr1}.

The strength of the tetrahedral ordering strongly affects the anomalous behavior of the liquids \cite{tetr1,tetr2}: the anomalies are
absent in the case of tin ($\lambda=18.5$) and carbon ($\lambda=26.2$), they are weak in the case of germanium ($\lambda=20$)
and strong in the case of liquid silicon ($\lambda=21.0$) and water ($\lambda=23.15$). One can see that the anomalies exist in a narrow
range of $\lambda$ and disappear if $\lambda$ is decreased or increased.

Basing on these results the dispersion curves of liquid silicon and water are simulated within the framework of SW model as
two examples of anomalous tetrahedral liquids. For both systems with calculate an equation of state and
dispersion curves of longitudinal and transverse excitations at an isochore with density anomaly, in order to
monitor the influence of the tetrahedrality on the anomaly of the dispersion curves in liquids.

\section{Systems and Methods}

In the present paper we two systems are considered: liquid silicon and water. Both of them are modeled by
the SW potential. The energy of a system with the SW potential is described by the following
equation \cite{sw}:

\begin{equation}\label{sw-pot}
E= \sum_{i} \sum_{j>i} \phi_2(r_{ij})+ \sum_{i} \sum_{j \neq i} \sum_{k>j} \phi_3(r_{ij}r_{ik} \theta_{ijk}),
\end{equation}
where

\begin{equation}
\phi_2(r_{ij})=A_{ij} \varepsilon_{ij} \left( B_{ij} ( \frac{\sigma_{ij}}{r_{ij}})^{p_{ij}} - (\frac{ \sigma_{ij}}{r_{ij}})^{q_{ij}} \right)exp(\frac{\sigma_{ij}}{r_{ij}-a_{ij}\sigma_{ij}}).
\end{equation}

\begin{equation}
\phi_3(r_{ij},r_{ik},\theta_{ijk})=\lambda_{ijk} \varepsilon_{ijk} \left( cos \theta_{ijk} - cos \theta_{0,ijk} \right)^2 exp(\frac{\gamma_{ij} \sigma_{ij}}{r_{ij}-a_{ij}\sigma_{ij}}) exp(\frac{\gamma_{ik} \sigma_{ik}}{r_{ik}-a_{ik}\sigma_{ik}}).
\end{equation}
The parameters of the potential for both silicon and water are given in Table I. The difference in the potentials for silicon and
water is in three parameters: depth of the potential well $\varepsilon$, effective size of the particles $\sigma$ and
parameter of "tetrahedrality" $\lambda$: higher $\lambda$ means higher range of tetrahedrality, i.e. water is more
tetrahedral then silicon.

\begin{table}
\begin{tabular}{|c|c|c|}
  \hline
    & Si & water  \\
 \hline
   $\varepsilon$ & 2.1683 & 0.268381 \\
  \hline
   $\sigma$ & 2.0951 & 2.3925 \\
   \hline
   a & 1.80 & 1.80 \\
   \hline
  $\lambda$ & 21.0 & 23.15 \\
   \hline
   $\gamma$ & 1.80 & 1.20 \\
   \hline
   $cos(\theta_0)$ & -1/3 & -1/3 \\
   \hline
   A & 7.05 & 7.05 \\
   \hline
   B & 0.602 & 0.602 \\
   \hline
   p & 4.0 & 4.0 \\
   \hline
   q & 0.0 & 0.0 \\

  \hline
\end{tabular}
\caption{coefficients SW potential for silicon and water Eq. \ref{sw-pot}. The parameters for silicon are taken from Ref. \cite{sw}
and the parameters for water from Ref. \cite{sw-water}.}
\end{table}

In both cases of liquid silicon and water a system of 8000 particles in a cubic box with
periodic boundary conditions is simulated by means of molecular dynamics method. The density of
silicon 2.503 $g/cm^3$ and the density of water is 0.99702 $g/cm^3$.
The selection of the density is based on the presence of the density anomaly in the system, i.e. negative values of
thermal expansion coefficient. Negative thermal expansion coefficient leads to appearance of a minimum in the dependence of
pressure on temperature along an isochor. This criterion in used in the present paper to prove the existence of the
density anomaly along the studied isochores.

In the case of liquid silicon the system is equilibrated for 2 ns in canonical ensemble (constant number of particles N,
volume V and temperature T). After that long micro-canonical simulation (constant number of particles N,
volume V and internal energy E) for more 5 ns is performed. During this stage the properties of the system are calculated. The time
step is $dt=0.0001$ ps. The temperatures vary from $T_{min}=1000$ K to $T_{max}=2000$ K. The temperature of minimum
pressure is $T_{DA}=1300$ K.

The simulation setup for water is similar to the one for liquid silicon. Time step is set to $dt=0.0005$ ps.
Equilibration period is 2 ns and the production period is 5 ns. Temperature varies from $T_{min}=220$ K to
$T_{max}=1000$ K. The melting point of the SW model of water is $T_m=274.6$ K \cite{sw-water}. The temperature of
the density anomaly at 1 atmosphere is $T_{DA}=250$ K \cite{sw-water}, i.e. it corresponds to the supercooled region.

Importantly, the SW model of water was fitted to reproduce the properties of water at low temperature. Because of this
at the temperatures above about 400 K the results are expected to be different from the ones of real water. However, the
goal of the present paper is to compare the qualitative behavior of the dispersion curves of two tetrahedral liquid
within the framework of the same model. Because of this so high temperatures are used in simulation, even if the model does not reproduce the experimental
results.

In order to find the excitation frequencies the velocity current autocorrelation functions are calculated. The longitudinal
and transverse parts of these functions are defined as:

\begin{equation}
C_L(k,t)=\frac{k^2}{N}\langle J_z({\bf k},t) \cdot J_z(-{\bf
k},0)\rangle
\end{equation}
and

\begin{equation}
C_T(k,t)=\frac{k^2}{2N} \langle J_x({\bf k},t)\cdot J_x(-{\bf
k},0)+J_y({\bf k},t) \cdot J_y(-{\bf k},0)\rangle
\end{equation}
where $J({\bf k},t)=\sum_{j=1}^N {\bf v}_j e^{-i{\bf k r}_j(t)}$
is the velocity current and wave vector $\bf{k}$ is directed
along the z axis \cite{hansenmcd,rap}. The dispersion curves of
longitudinal and transverse excitations can be obtained as the
location of maxima of Fourier transforms $\tilde{C}_L({\bf
k},\omega)$ and $\tilde{C}_T({\bf k},\omega)$ respectively.

\section{Results and Discussion}

\subsection{Liquid silicon}

\begin{figure}
\includegraphics[width=6cm,height=6cm]{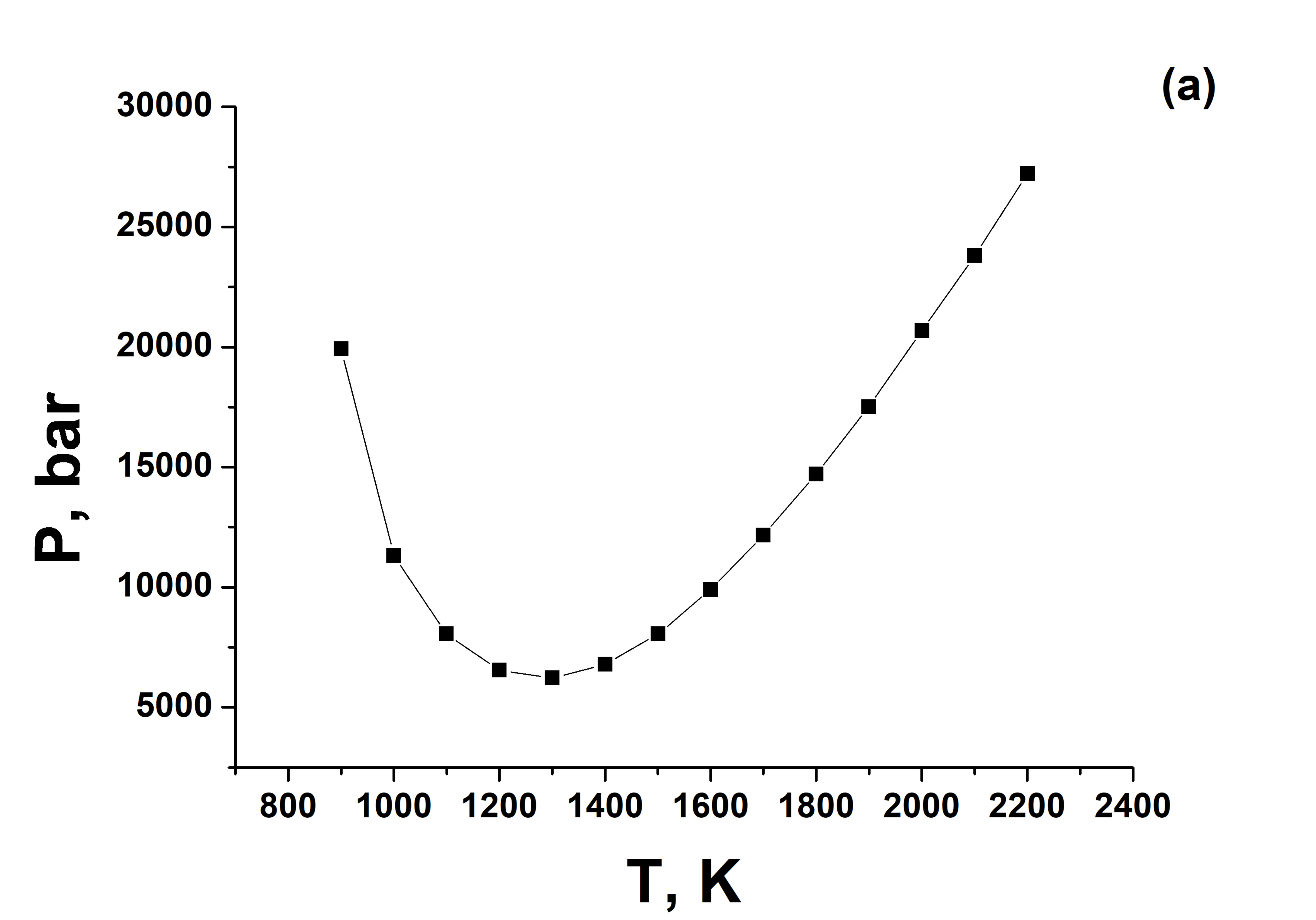}%

\includegraphics[width=6cm,height=6cm]{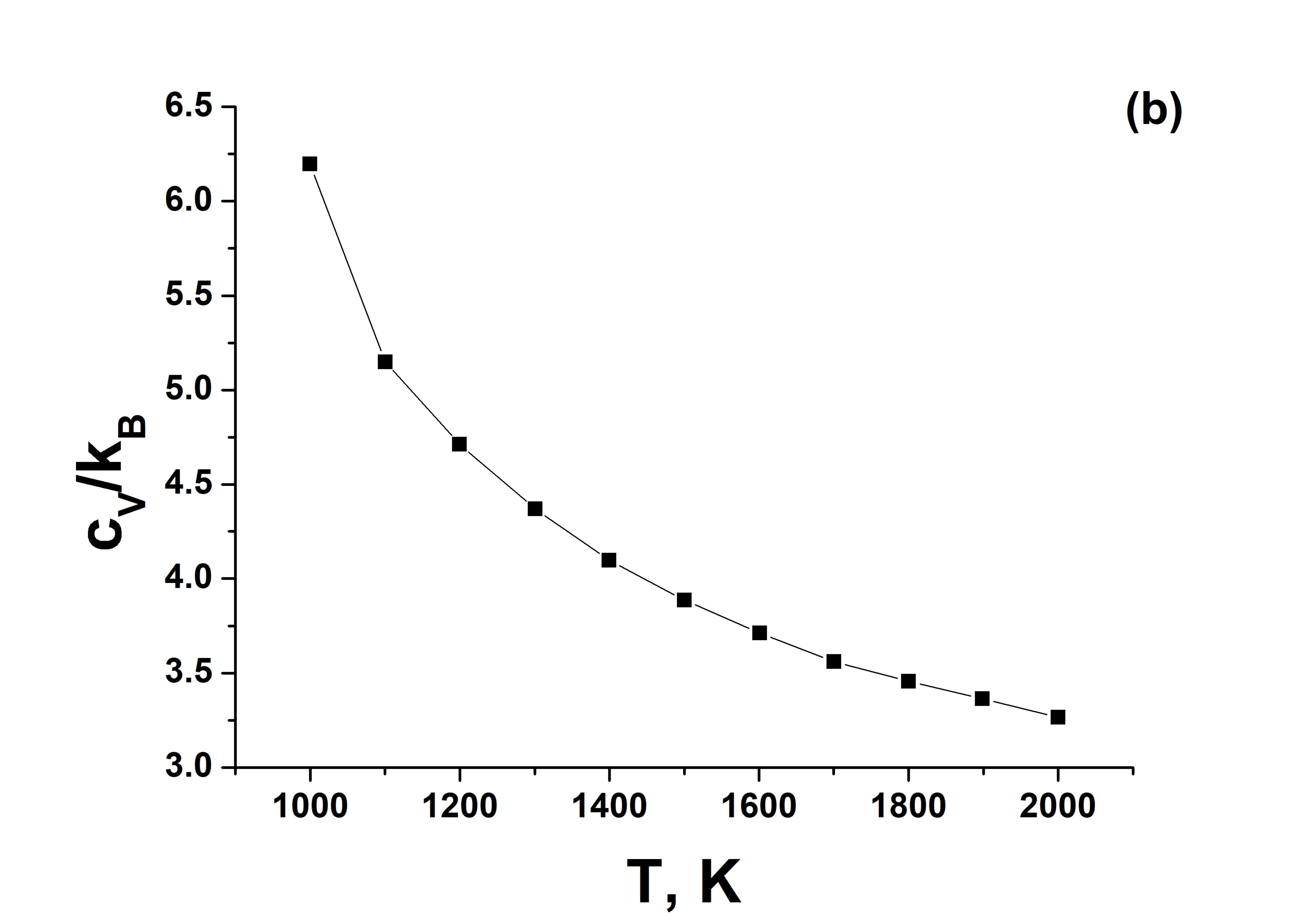}%

\caption{\label{eossi} (a) The equation of state of liquid silicon along the isochore $\rho=2.503$ $g/cm^3$.
(b) The isochoric heat capacity of silicon at the same density.}
\end{figure}

\begin{figure}
\includegraphics[width=6cm,height=6cm]{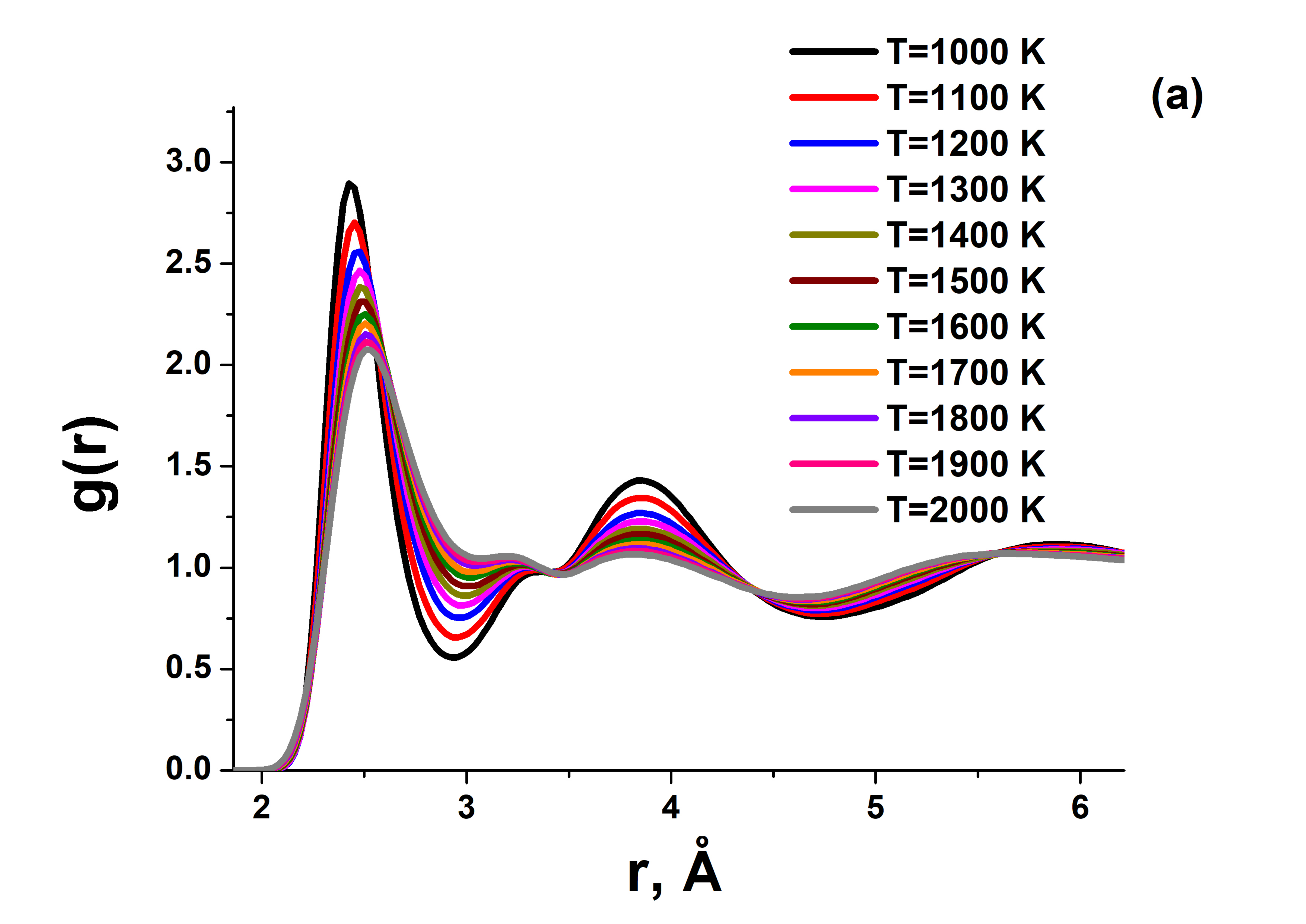}%

\includegraphics[width=6cm,height=6cm]{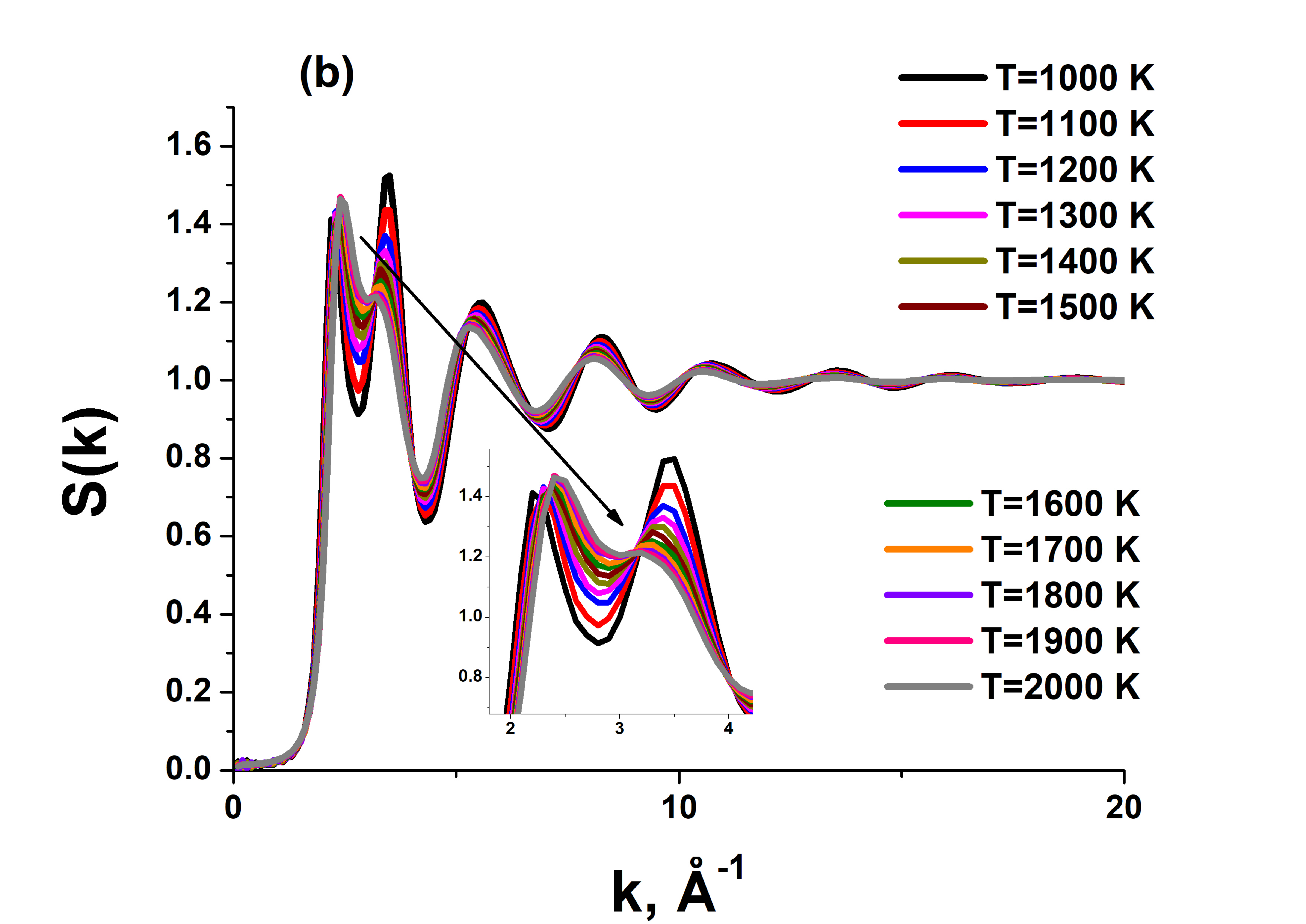}%

\caption{\label{rdf-si} (a) The radial distribution functions of liquid silicon along the isochore $\rho=2.503$ $g/cm^3$.
(b) The structure factors of liquid silicon at the same density.}
\end{figure}

We first describe results for the case of liquid silicon. Fig. \ref{eossi} shows the equation of state
along isochore $\rho=2.503$ $g/cm^3$ which is used in the present work. One can see a minimum at
$T_{DA}=1300$ K, which confirms the presence of the density anomaly in the system, in agreement with previous
works \cite{dasi1,dasi2,dasi3,dasi4}. Moreover, the isochoric heat capacity of liquid silicon has very high values at low temperature (Fig. \ref{eossi} (b)).
Such behavior appears when liquid demonstrates a smooth structural crossover \cite{cvlarge,cvlarge1}. This crossover can be seen from
radial distribution functions (RDFs) and structure factors of the system shown in Fig. \ref{rdf-si}. Both RDFs and structure factors have complex shape.
In the case of the RDFs the second peak has a shoulder at lower values of $r$, which means that the second coordination shell is scattered. This
scattering should be responsible for the appearance of the density anomaly.

The complex nature of the local structure of liquid silicon is even more pronounced in structure factors (Fig. \ref{rdf-si} (b)).
Structure factors demonstrate two clear peaks. Al low temperatures the second peak appears to be higher then the first. However, when
the temperature increases the first peak goes up, while the second one goes down. Therefore, smooth change of the local structure is observed.

Basing on the presence of water-like anomalous behavior and smooth structural crossover the question is whether liquid silicon
demonstrates anomalous dependence of the dispersion curves similar to the one observed in a model core-softened system
\cite{disp-anom-wat} and SPC/E model of water \cite{disp-anom-wat,disp-anom-135}.

\begin{figure}
\includegraphics[width=6cm,height=6cm]{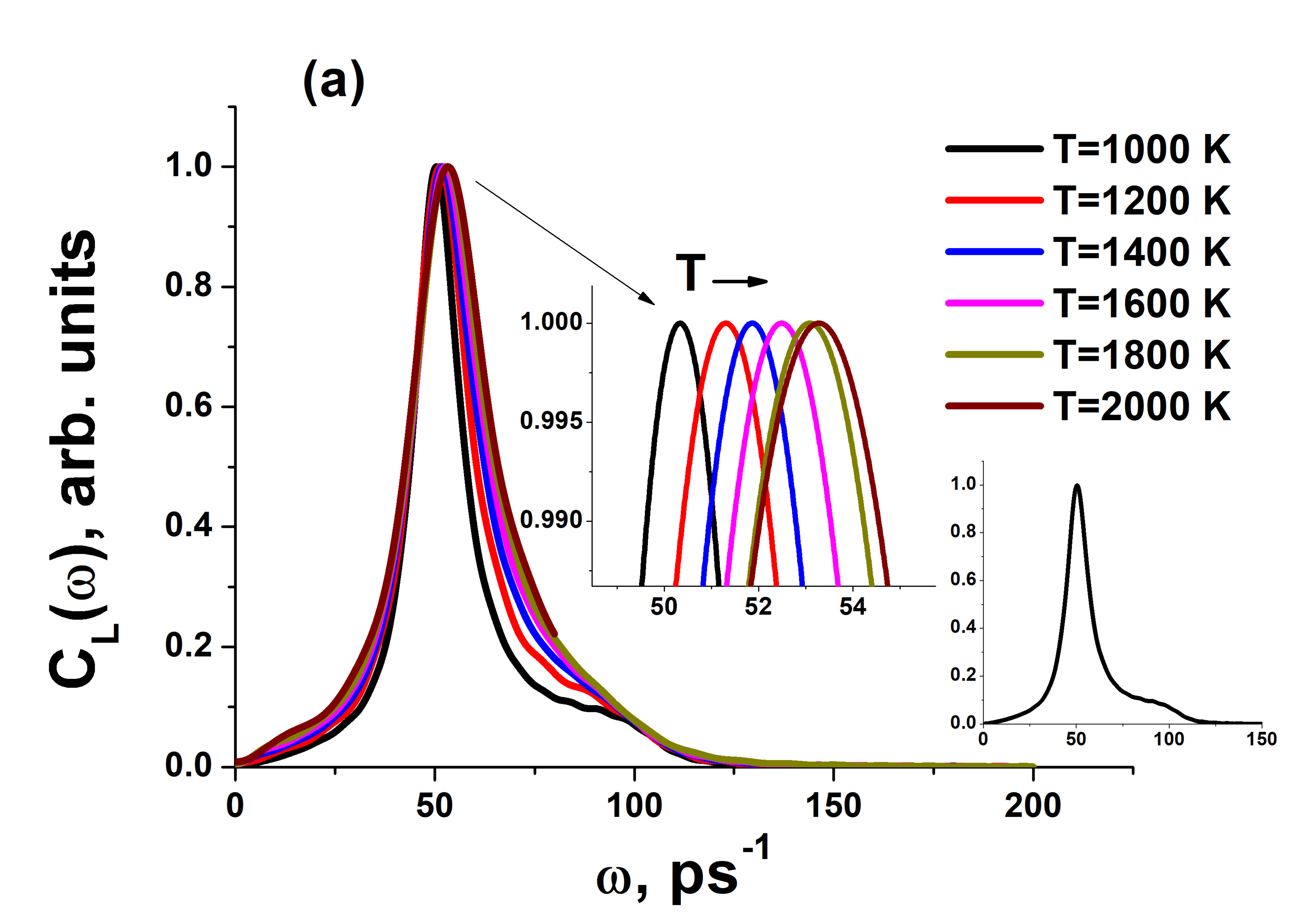}%

\includegraphics[width=6cm,height=6cm]{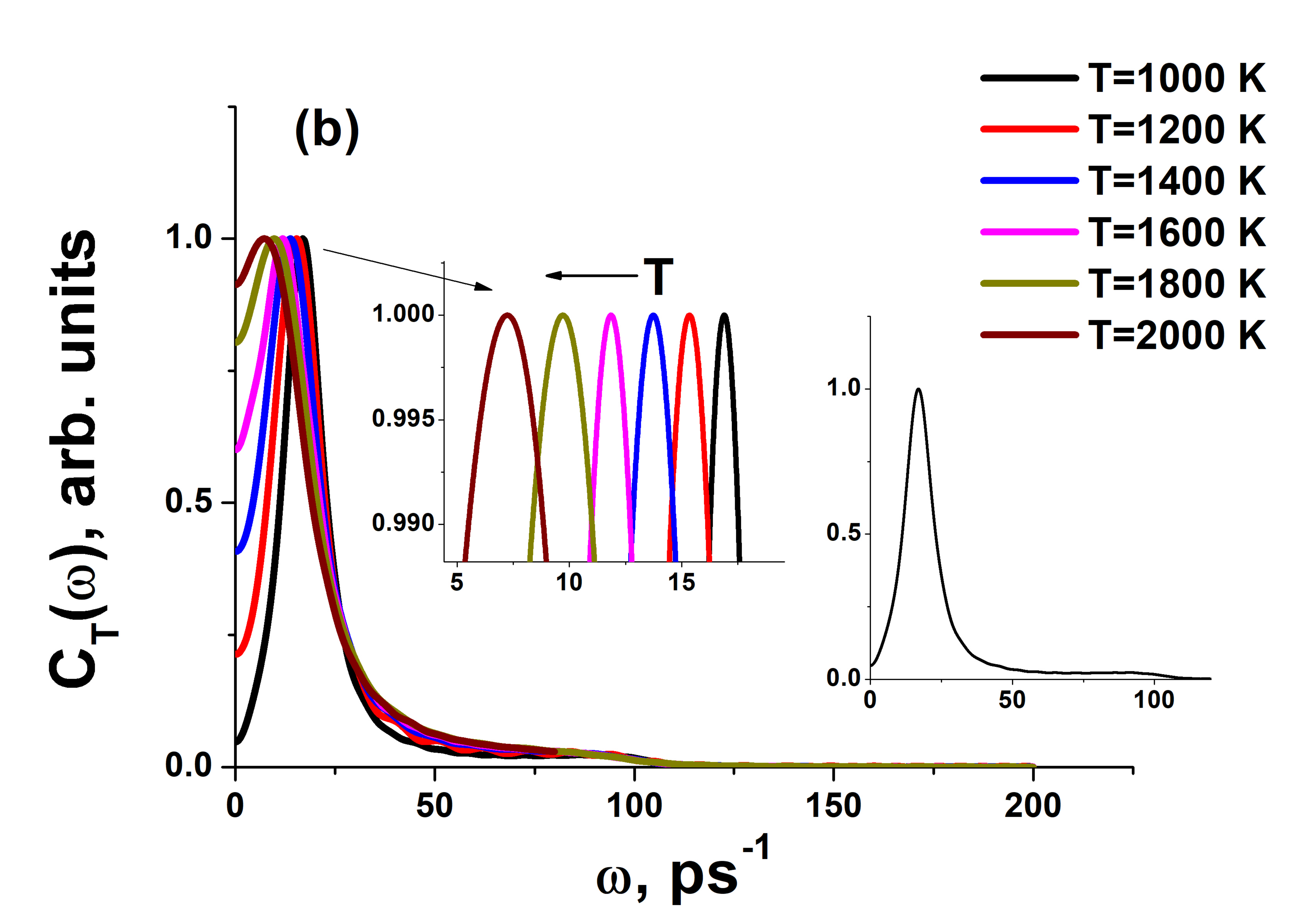}%

\caption{\label{wlt-si} (a) Examples of Fourier image of longitudinal current autocorrelation functions of liquid silicon
at $k=0.592$ $\AA$ at different temperatures. The left inset enlarges the region of maxima of the curves. The arrow indicates the
direction of increasing of the temperature. The right inset enlarges the curve for $T=1000$ K. (b) The same for the transverse excitations.}
\end{figure}

Fig. \ref{wlt-si} (a) and (b) show the examples of Fourier image of longitudinal and transverse excitations of liquid silicon.
Unlike the case of a simple liquid, these Fourier images cannot be approximated by a simple peak, since they also demonstrate a shoulder
at higher frequencies. This is consistent with the results for water obtained in Refs. \cite{3peaks,ramil} In the present paper we consider the frequencies of
the main peak only as it is the most powerful process in the system.

The location of maxima of the curves shown in Fig. \ref{wlt-si} (a) and (b) follows normal regime: the frequency of the maximum increases
with temperature for longitudinal excitations and decreases with a temperature increase for the transverse ones. However, the frequency shift
of longitudinal excitations is extremely weak: while the temperature changes two-fold (from $T=1000$ K to $T=2000$ K) the shift of
frequency is just about $4 \%$.

\begin{figure}
\includegraphics[width=6cm,height=6cm]{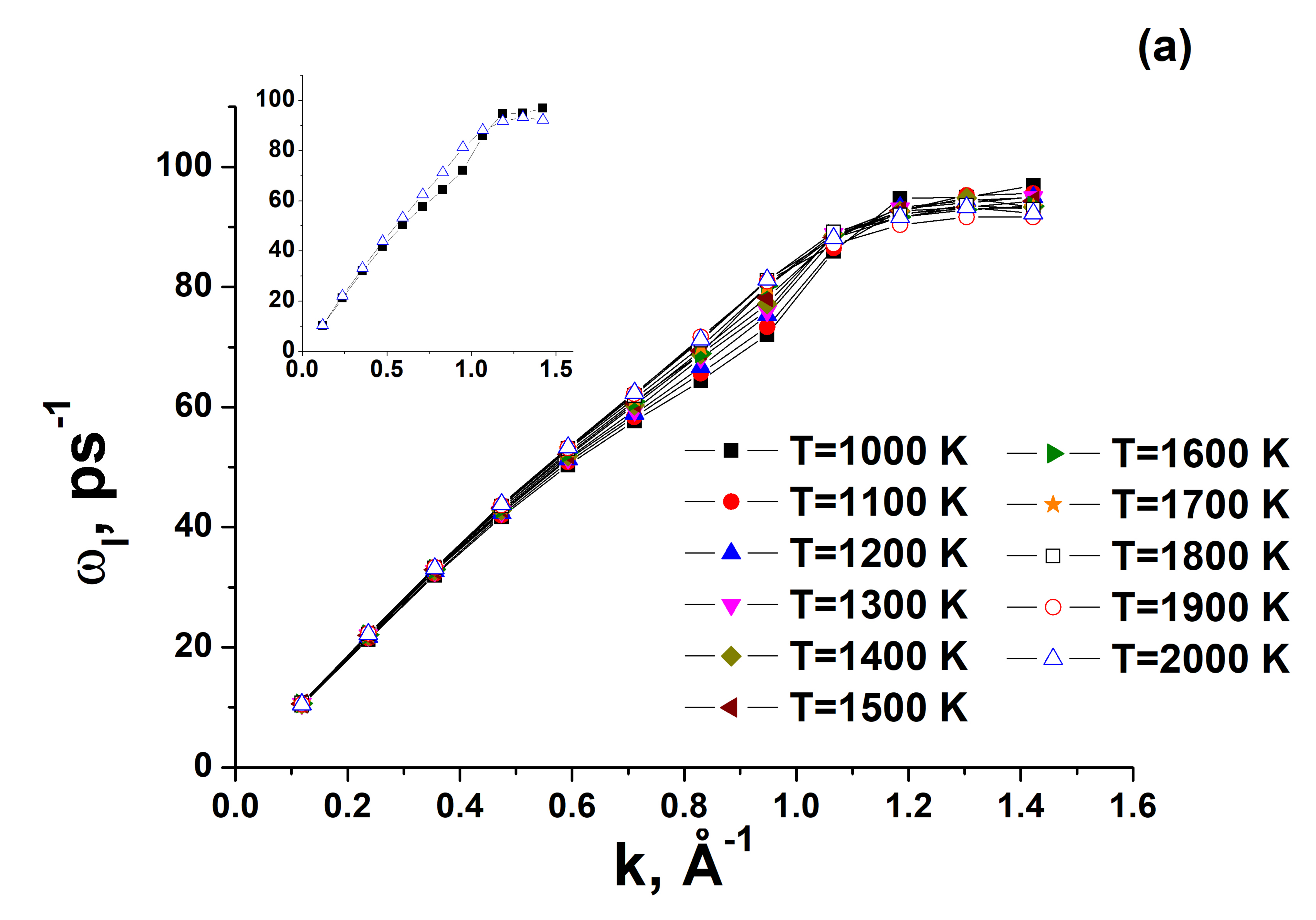}%

\includegraphics[width=6cm,height=6cm]{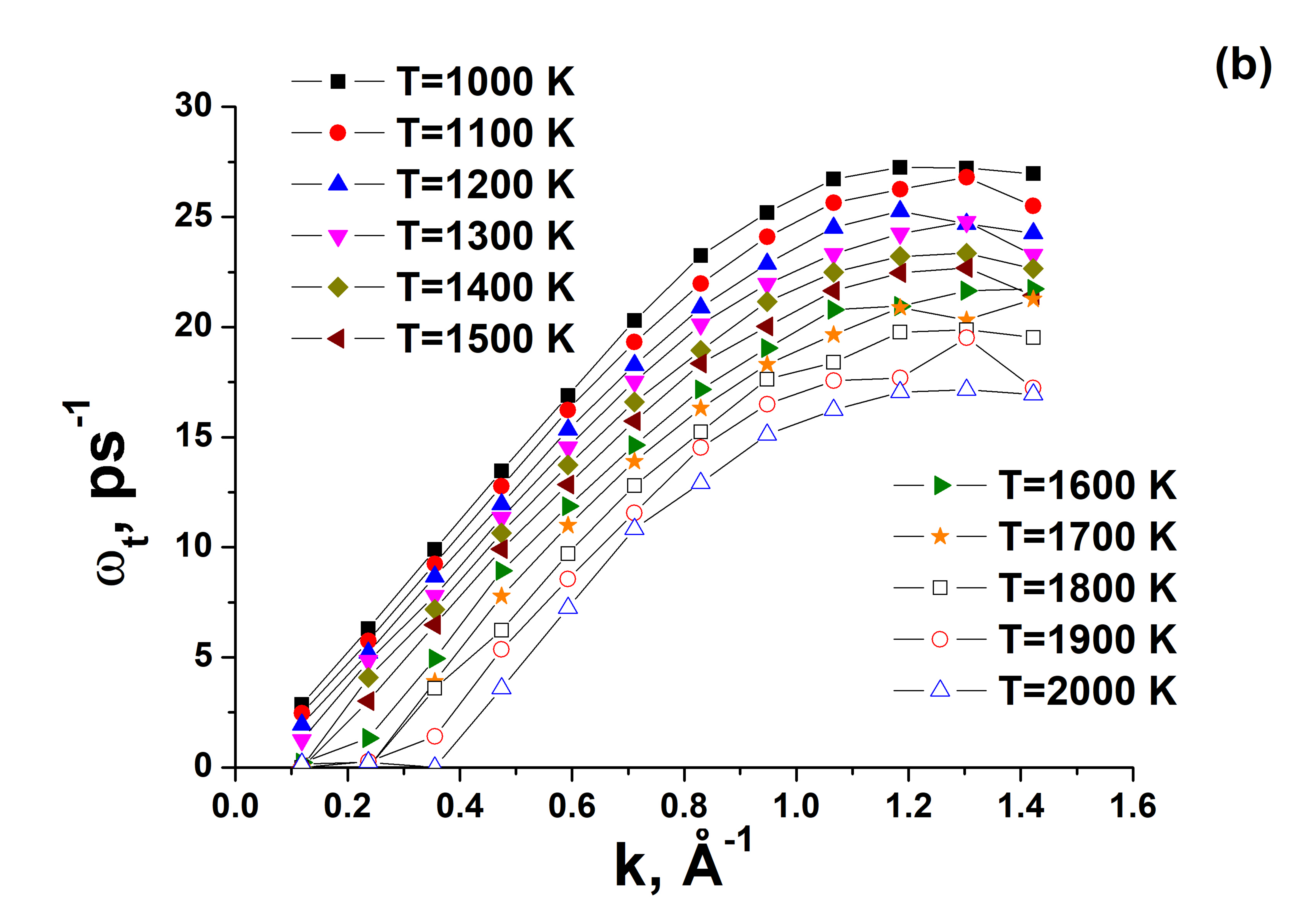}%

\caption{\label{disp-si} The dispersion curves of (a) longitudinal and (b) transverse excitations of liquid silicon at $\rho=2.503$ $g/cm^3$ and
different temperatures. The inset on the panel (a) gives a comparison of the dispersion curves at the lowest and highest temperatures studied
($T=1000$ K and $2000$ K).}
\end{figure}

Fig. \ref{disp-si} (a) and (b) show the dispersion of longitudinal and transverse excitations of liquid silicon. One can see that
longitudinal excitations demonstrate very modest temperature dependence. The inset of the panel (a) gives a comparison
of the dispersion curves at the lowest and the highest temperatures ($T=1000$ K and $2000$ K respectively). One can see that the
difference between these curves does not exceed $4 \%$ which can be attributed to an error of calculations.

Importantly, in the case of a simple liquid the excitation frequency should increase with temperature. In this respect
extremely weak temperature dependence of the excitation frequencies of liquid silicon should be considered as
anomaly, even if it is not so pronounced as in RSS and SPC/E water, where decrease of the frequency with a temperature increase was observed
\cite{disp-anom-wat,disp-anom-135}.

The dispersion curves of transverse excitations behaves in an usual way: the frequency at some fixed wave vector ${\bf k}$ decreases
with temperature. At the lowest temperatures no band gap is observed. It appears at $T=1700$ K. The width of the band gap
increases with temperature. The width of the band gap becomes equal to the width of the first Brillouin zone at the Frenkel line \cite{fr1,fr2,fr3,fr4,fr5,fr6}.
From Fig. \ref{disp-si} it can be concluded that the temperatures of the present study are well below the Frenkel temperature of liquid silicon.

\subsection{Water}

\begin{figure}
\includegraphics[width=6cm,height=6cm]{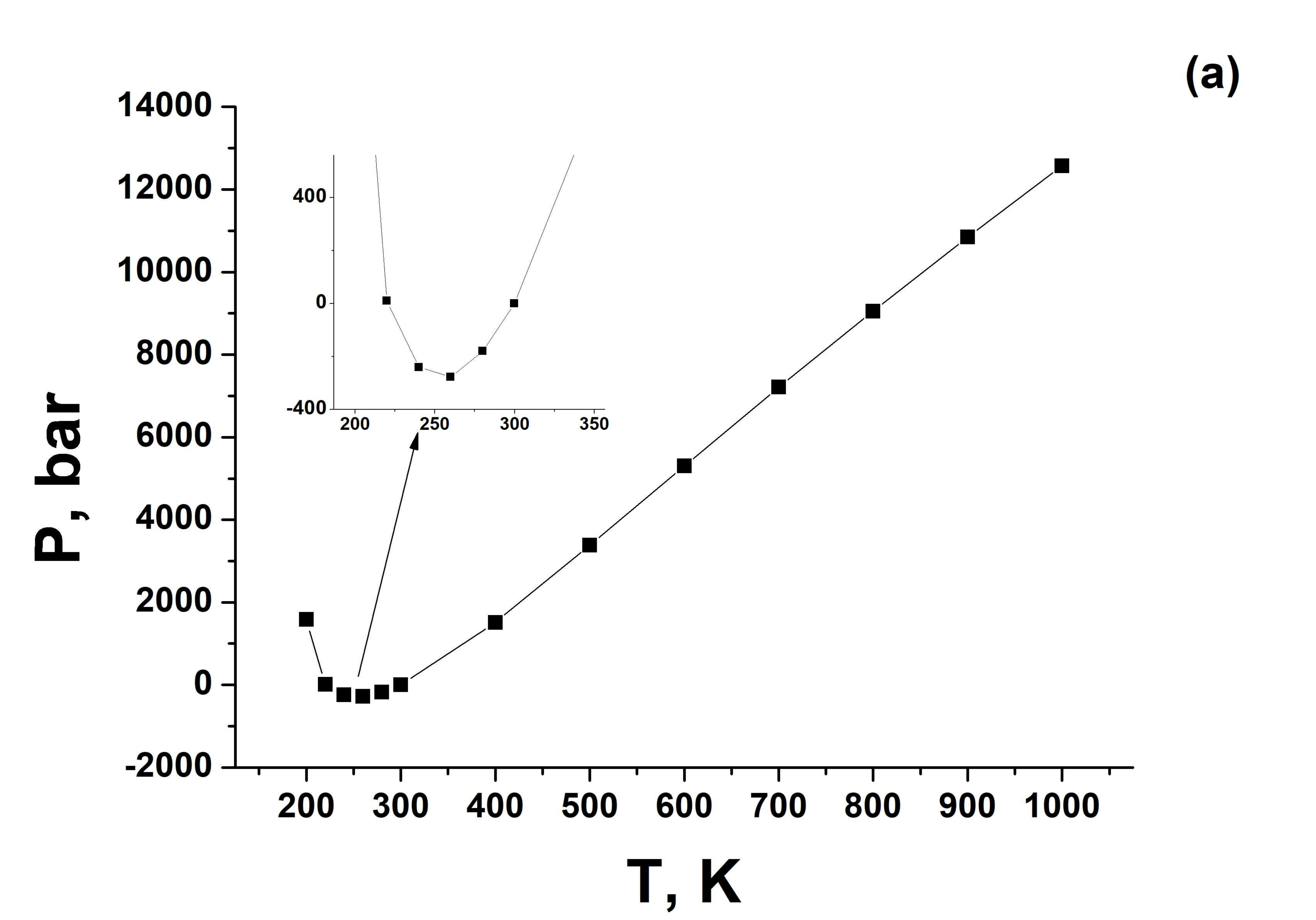}%

\includegraphics[width=6cm,height=6cm]{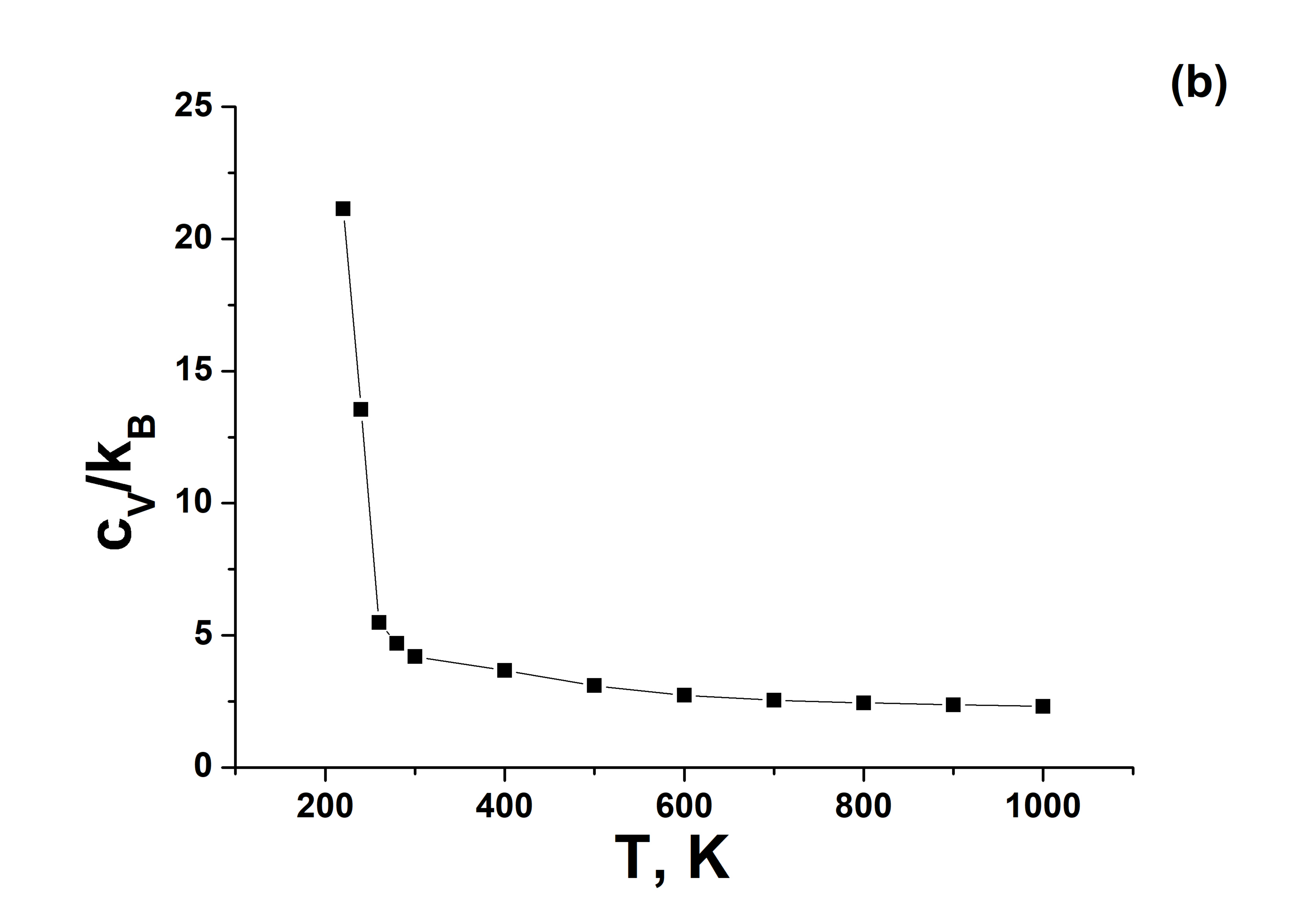}%

\caption{\label{eos-wat}  (a) The equation of state of water along the isochore $\rho=0.997$ $g/cm^3$.
(b) The isochoric heat capacity of water at the same density.}
\end{figure}

The same analysis was performed for water. Fig. \ref{eos-wat} (a) and (b) show the equation of state and the isochoric heat capacity
of water along isochor $\rho=0.997$. The equation of state demonstrates a minimum, i.e. the density anomaly. The
isochoric heat capacity has very large value, i.e. smooth structural crossover takes place. This is very similar to the case of
liquid silicon described above and to the behavior of the SPC/E water model \cite{disp-anom-wat,disp-anom-135}.

\begin{figure}
\includegraphics[width=6cm,height=6cm]{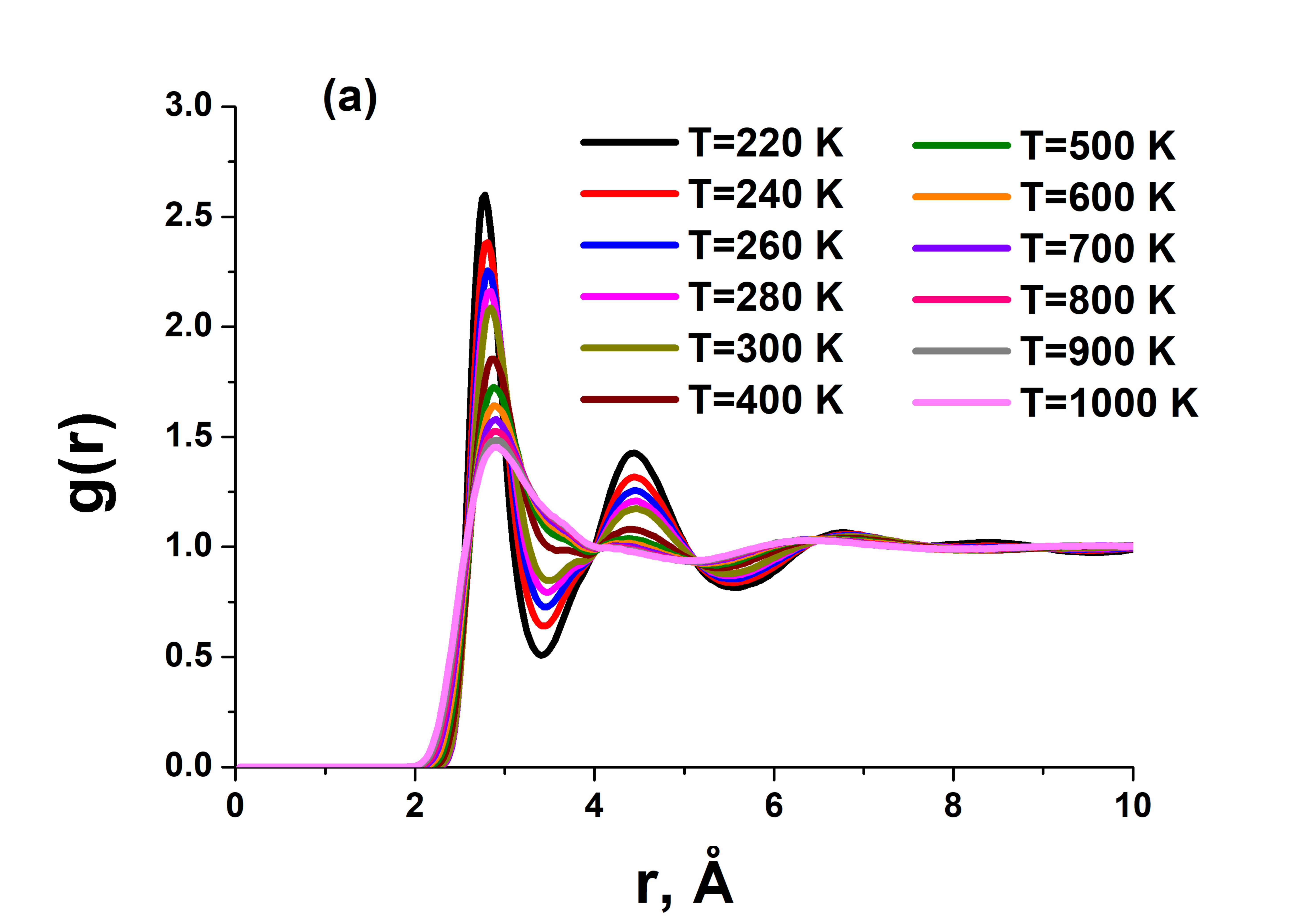}%

\includegraphics[width=6cm,height=6cm]{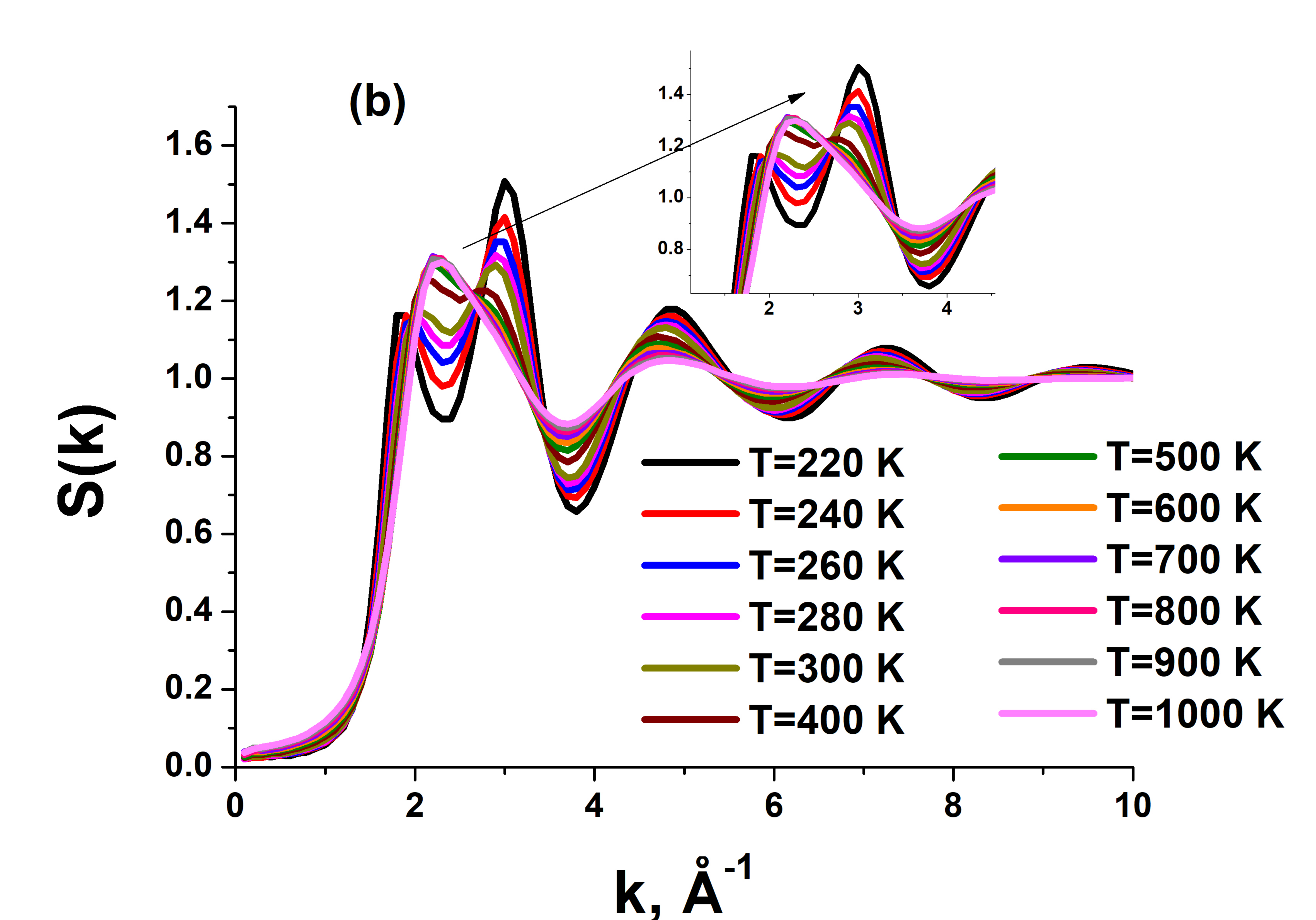}%

\caption{\label{rdf-wat} (a) The radial distribution functions of water at the isochore $\rho=0.997$ $g/cm^3$.
(b) The structure factors of water at the same density.}
\end{figure}

Fig. \ref{rdf-wat} (a) and (b) give the RDFs and structure factors of water. One can see that the structure of water changes rapidly
on heating. While at low temperatures two clear peaks are observed, at high ones only the first peak with wide shoulder from right is
present. The same effect is seen in structure factors: two peaks at low temperatures and a single peak at high ones.

\begin{figure}
\includegraphics[width=6cm,height=6cm]{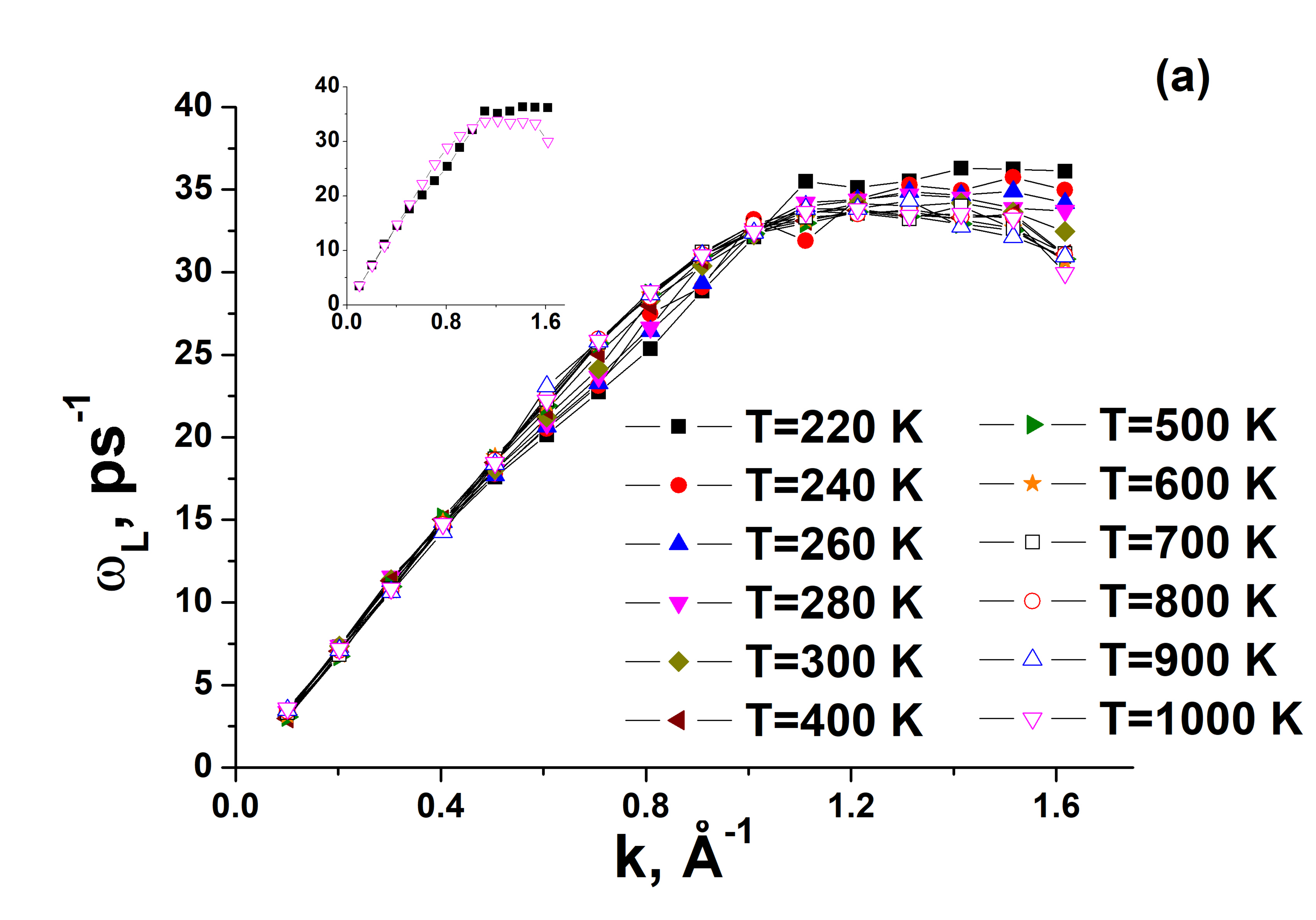}%

\includegraphics[width=6cm,height=6cm]{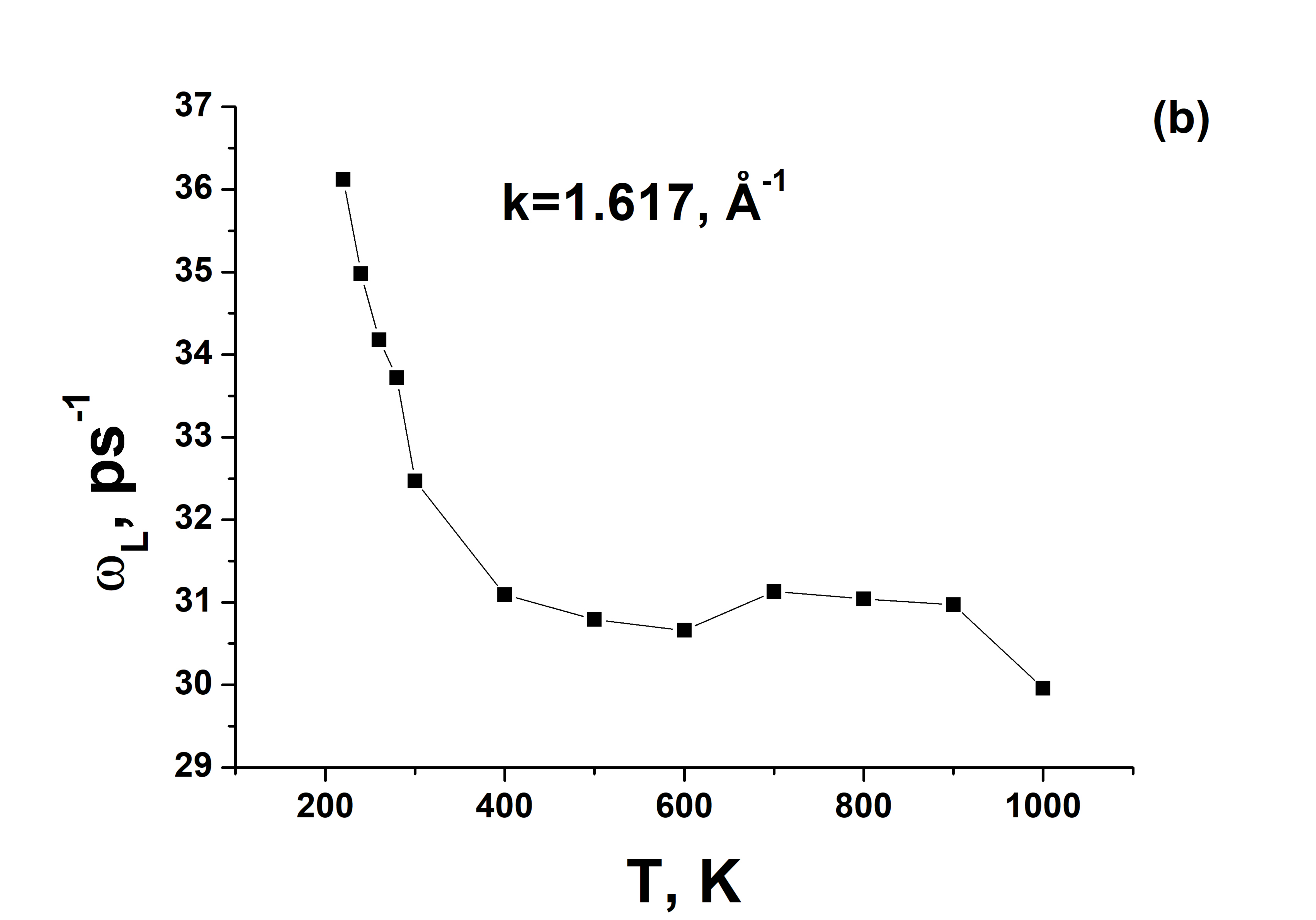}%

\includegraphics[width=6cm,height=6cm]{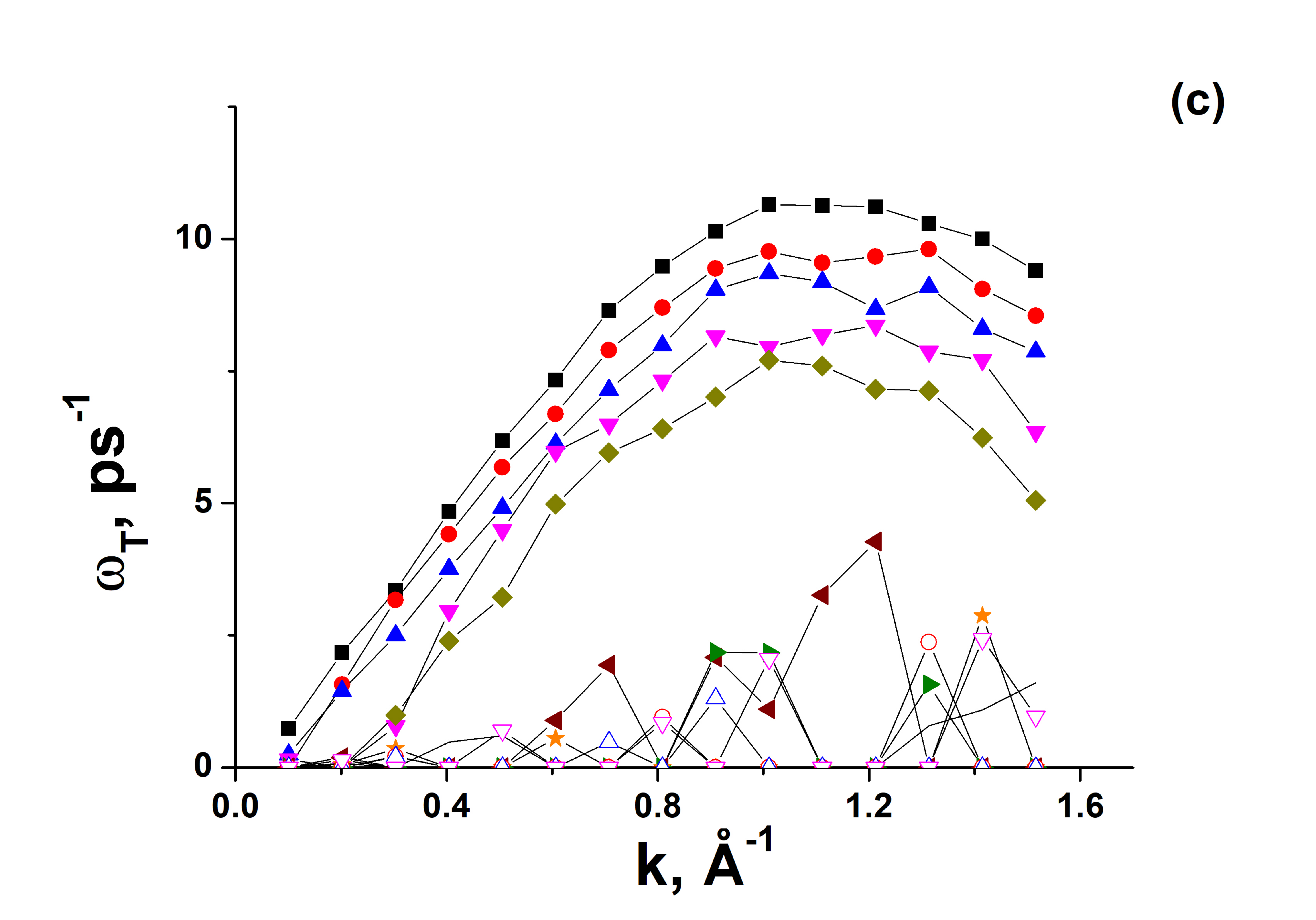}%

\caption{\label{disp-wat} (a) The dispersion curves of longitudinal excitations of water at $\rho=0.997$ $g/cm^3$ and
different temperatures. The inset on the panel (a) gives a comparison of the dispersion curves at the lowest and highest temperatures studied.
(b) The temperature dependence of the frequency of longitudinal excitations of water at $k=1.617$ $\AA ^{-1}$. (c) The dispersion curves of transverse excitations of water at the same density and the same set of temperatures. The notation of the curves of the panel (c) is the same that of the panel (a).}
\end{figure}

Fig. \ref{disp-wat} (a) shows the dispersion curves of longitudinal excitations in water at $\rho=0.997$ and a set of temperatures.
One can see that at moderate wave vectors the temperature dependence of the frequency is like in a simple liquid, i.e.
the frequency increases with temperature. However, at $k=1$ $\AA^{-1}$ all the curves cross and the temperature dependence of
the frequency becomes anomalous. Fig. \ref{disp-wat} (b) demonstrates the frequency of longitudinal waves at $k=1.617$ $\AA^{-1}$.
One can see that the frequency slightly decreases with a temperature increase. There is a small bump in the temperature range 600 - 900 K, which,
however, does not change the general trend of the curve. Therefore, in the case of water the anomalous temperature dependence
of the longitudinal excitation frequency does take place.

The transverse excitations of water behave as in a simple liquid, i.e. the frequency decreases with a temperature increase.
Moreover, one can see unambiguous transverse excitations at the temperatures from $T=220$ K to $T=300$ K. At $T=400$ K
we do not observe well recognized transverse excitations. Therefore, the Frenkel temperature of the SW water is somewhere between 300 and 400 K.

The SW model of water was fitted to reproduce the behavior of water in the vicinity of the melting point. Because of this the model does not
reproduce the behavior of water at high temperatures used in the present study. This is clearly seen from the estimation of the Frenkel temperature
of the SW model which appears to be between 300 and 400 K. In the case of the SPC/E model of water the Frenkel temperature is 460 K \cite{fr-wat}, while
for the TIP4P/2005 model it is $520$ K \cite{kostya-china}. However, it is important to study the dispersion curves of SW-water for the sake of
comparison with the ones of liquid silicon and to show that the same model can demonstrate both normal or anomalous behavior of the dispersion curves
depending on the parameters of the potential.

\bigskip

The behavior of different anomalies (density anomaly, diffusion anomaly and structural anomaly) of a model core-softened system
with different parameters of interaction potential (Repulsive Shoulder System, RSS) was studied in our previous works \cite{s135,s135b,s135c,s135d,s135e}.
The behavior of the system is determined by the width of the repulsive shoulder of the potential.
It was shown that changing the width of the shoulder can lead to disappearance of some of the anomalies \cite{s135d,s135e}.
Importantly, all studied systems demonstrated negative slope of the melting line and the location of anomalies was at the densities higher then the
melting curve. When the width of the shoulder becomes larger the melting line and the regions of all anomalies (density, diffusion and structure) move to
smaller densities. However, the "rate" of this motion is different for the different curves (the melting line
and the boundaries of the anomalous regions). As a result some of the anomalies rapidly go below the melting line
and disappear. We expect that similar phenomenon can take place in the case of the anomaly of the longitudinal excitations in the SW systems: changing
the parameters of the potential makes the anomaly more pronounced in the case of water and less pronounced in the case of liquid silicon. Moreover,
this can be the reason of disappearance of other anomalies, when the tetrahedrality parameter $\lambda$ is changed. However, verification of this
assumption requires determination of the phase diagram of the systems with different $\lambda$ which is out of the scope of the present paper.

\section{Conclusions}

In conclusion, we have examined the dispersion curves of both longitudinal and transverse excitations of liquid silicon and water in frames of SW potential models
fitted for these substances. As it was shown before, the SPC/E model of water demonstrates anomalous temperature dependence of the frequency of
longitudinal excitations: while in a simple liquid the frequency increases with temperature, in SPC/E model of water in can decrease. The same effect
was found in the SW model of water. In the case of liquid silicon we observe extremely weak temperature dependence of the longitudinal frequency, i.e.
the anomaly is extremely weak. The behavior of transverse excitations is qualitatively similar in both liquid silicon and water: at fixed
$k-$vector the frequency decreases with a temperature increase. At the same time the Frenkel temperature of liquid silicon appears to be very high (more then
$T=2000$ K, which is the highest temperature of the present study), while in water the Frenkel temperature is somewhere between $300$ and $400$ K, i.e.
very close to the melting point.

\bigskip

This work was carried out using computing resources of the federal
collective usage center "Complex for simulation and data
processing for mega-science facilities" at NRC "Kurchatov
Institute", http://ckp.nrcki.ru, and supercomputers at Joint
Supercomputer Center of the Russian Academy of Sciences (JSCC
RAS). The work was supported by the Russian Science
Foundation (Grants No 19-12-00111).

\end{document}